\documentclass[%
 reprint,
 amsmath,amssymb,
 aps,
]{revtex4-2}

\usepackage{url}
\usepackage{graphicx}
\usepackage{dcolumn}
\usepackage{bm}
\usepackage{xcolor,soul,framed} 
\usepackage{subfigure}
\usepackage{fancyhdr}
\usepackage{booktabs}
\usepackage{tabularx}
\usepackage{comment}
\usepackage{blindtext}

\def\andname{\hspace*{-0.5em},}

\begin{document}



\widetext

\pagenumbering{gobble}

\noindent \textbf{Copyright:}
\copyright 2025  by American Physical Society. All rights reserved.\\

\noindent Individual articles are copyrighted by the APS, as indicated on each article.\\

\noindent Individual articles may be downloaded for personal use; users are forbidden to reproduce, republish, redistribute, or resell any materials from this journal in either machine-readable form or any other form without permission of the APS or payment of the appropriate royalty for reuse. \\

\noindent \textbf{Disclaimer:} This work has been published in \textit{Physical Review Applied}. \\

\noindent Citation information: DOI 10.1103/85sy-qbk6

\pagenumbering{arabic}

\title{{Kapitza-Inspired Stabilization of Non-Foster Circuits via Time Modulations}}


\author{Antonio Alex-Amor}
\thanks{Antonio Alex-Amor is currently affiliated with the Dept. of Electronics and Communications Technology, RFCAS Group, Universidad Autónoma de Madrid, 28049 Madrid, Spain.}
\author{\def\andname{,}Grigorii Ptitcyn, Nader Engheta}
\thanks{Contact author: engheta@seas.upenn.edu}
\affiliation{
 University of Pennsylvania, Department of Electrical and Systems Engineering, Philadelphia, Pennsylvania 19104, United States
}%
%

\begin{abstract}
With his formal analysis in 1951, the physicist Pyotr Kapitza demonstrated that an inverted pendulum with an externally vibrating base can be stable in its upper position, thus overcoming the force of gravity. Kapitza's work is an example that an originally unstable system can become stable after a minor perturbation of its properties or initial conditions is applied. Inspired by his ideas, we show how non-Foster circuits can be stabilized with the application of external \textit{electrical vibration}, i.e., time modulations. Non-Foster circuits are highly appreciated in the engineering community, since their bandwidth characteristics are not limited by passive-circuits bounds. Unfortunately, non-Foster circuits are usually unstable and they must be stabilized prior to operation. Here, we focus on the study of non-Foster $L(t)C$ circuits with time-varying inductors and time-invariant negative capacitors. We find an intrinsic connection between Kapitza's inverted pendulum and non-Foster $L(t)C$ resonators. Moreover, we show how positive time-varying modulations of $L(t)>0$ can overcome and stabilize  non-Foster negative capacitances $C<0$. These findings open up an alternative manner of stabilizing electric circuits with the use of time modulations, and lay the groundwork for application of, what we coin \textit{Vibrational Electromagnetics}, in more complex media.   
\end{abstract}

\maketitle


\section{\label{sec:introduction}Introduction}

 In 1951, Nobel laureate Pyotr Kapitza introduced a theory that effectively described the physics of one of the systems that physicists were unable to interpret at that time: the inverted pendulum with a vibrating base \cite{Kapitza1, Kapitza2}. The history of this problem dates back to 1908, when physicist Andrew Stephenson suggested that the upper position in a conventional pendulum could become stable with the addition of a high-frequency vibrating base with a small amplitude of vibration \cite{Stephenson1908}. The addition of this external element can indeed counteract the force of gravity that pulls the pendulum downwards, causing the pendulum to swing around its upper position without falling down.  

Kapitza's approach was based on the separation of the time scales and the use of time-averaging techniques \cite{Butikov2017KapitzaS}. He assumed that the external vibrating base oscillates with a frequency $\Omega$ much higher compared to the natural oscillating frequency of the pendulum, $\omega_0 \equiv \sqrt{g/l}$ ($g \approx 9.81$ m/s$^2$ is the acceleration of gravity and $l$ is the length of the pendulum). If, additionally, the amplitude of the modulation $\delta$ is small compared to the length of the pendulum $l$, \emph{perturbation} techniques can be invoked to reach an approximate analytical solution.  In a scenario with these characteristics, it is possible to describe the motion (i.e., the pendulum's angle $\theta(t)$ in Fig.~\ref{fig1}) of the inverted pendulum with a vibrating base by separating the total contribution into two main terms: the \emph{slow and fast components} of $\theta(t)$ = $\theta_s(t)$ + $\theta_f(t)$ . The slow component $\theta_s(t)$, a low-frequency term, represents the main contribution. On the other hand,  the fast component $\theta_f(t)$, a small-amplitude and high-frequency term, is a secondary perturbative contribution.  See the supplementary materials for some details \cite{SuppMaterial}.

Kapitza's ideas have transcended the scientific community beyond their initial application, leading to the subject of \textit{Vibrational Mechanics}. A variety of stimulations with different time scales are commonly seen in different fields of science and engineering. The overall low-frequency response of linear and, especially, nonlinear systems excited by multiple inputs may be significantly affected by weak time-varying high-frequency contributions \cite{Yang2024, Blekhman2000, Blekhman2004}. This phenomenon is named as \emph{Vibrational Resonance}, term originally coined by Landa and McClintock \cite{Landa2000} when studying high-frequency nonlinear oscillators and their connection to stochastic terms. Kapitza-inspired techniques have been applied to the analysis of a wide variety of physical, electronic and biological systems, such as the Chua's circuit \cite{Chua2013}, lasers \cite{Chizhevsky2021}, logic gates \cite{Murali2021}, energy harvesters \cite{Zhang2022},   quantum wells \cite{Olusola2020} {and critical points \cite{KapitzaQuantum2024}}, Hindmarsh-Rose neuronal systems \cite{Wang2021}, stratified media \cite{Rizza2013} or time-periodic potentials \cite{Muniz19, Alberucci2022}. 

Concurrently, the field of metamaterials has recently gained renewed attention with the inclusion of \emph{temporal modulations} as part of their inner structure \cite{caloz2019spacetime, galiffi2022photonics}. The so-called space-time or four-dimensional (4D) metamaterials are engineered structures whose mechanical or electrical properties change over space and time \cite{engheta2023four}. The use of time in material parameter variation as a new degree of freedom has allowed for innovative designs that overcome the performance of their static, time-invariant counterparts, such as magnetless non-reciprocal devices \cite{estep2014magnetic, wu2019isolating}, beamformers and beamsteerers \cite{taravati2022microwave, Moreno23spacetime}, frequency converters \cite{taravati2018aperiodic, Li2023prism}, power combiners \cite{Wang2021power}, or temporal lenses \cite{pacheco2023temporal}. 

In a different context, temporal modulations have enabled an alternative scheme to produce non-Foster electronic components \cite{Hrabar2020, Hrabar2022, Grigorii2022_conference, Grigorii2023}; namely, circuit elements that do not satisfy Foster's reactance theorem \cite{Foster1924}. Non-Foster circuits are active structures whose bandwidth characteristics are not limited by Bode-Fano and Carlin-La Rosa bounds \cite{Bode1945, Fano1948, Carlin1953}. For instance, antenna and metamaterial communities have benefited of their use to improve impedance matching performance \cite{Sussman2009}. 

{The first patented ideas on the production of non-Foster elements date back to the 1920s and 1930s \cite{Latour1923, Dolmage1927, Mathes1927}. At present, the production of non-Foster circuit components, such as negative resistors, inductors or capacitors, lies on well-established technology, thus the ideas presented here can, in principle, be put into practice.  One of the conventional methods for implementing  non-Foster circuit components  is  the use of negative impedance converters  \cite{Linvill1953}. More specifically, it is common to design feedback schemes  involving cross-coupled transistors to create non-Foster components such as negative capacitors \cite{Ghadiri2014, Covington2014}. The circuit topology in \cite{Covington2014} may serve as a reference:
two main transistors (plus four additional ones acting as current sources),  where the gate of each of the main transistors
is connected to the drain of the other, thus creating a negative capacitance.    }

{Despite the advantages they offer in engineering, non-Foster components are usually unstable and their stabilization is a challenging task \cite{Brownlie1966, Stearns2012, Tofigh2020}.} In the present work, inspired by Kapitza's ideas and the field of Vibrational Mechanics, we analyze the stability conditions of time-varying non-Foster resonators. First, we will focus on scenarios where the inductor is time modulated and the capacitor  has a time-invariant negative capacitance. Then, we will show how a time modulated \emph{positive} inductance $L(t)>0$ can stabilize a negative element capacitance $C<0$. This phenomenon, which may seem unexpected or counterintuitive at first, is a result of a vibrational resonance occurred in the non-Foster resonator. Moreover, we will show that the frequency of the already stabilized non-Foster resonator can be tuned over time by dynamically changing $L(t)$.

\section{Connection between the Non-Foster $L(t)C$ and Kapitza's Inverted Pendulum}

\begin{figure}[!t]
    \centering
\subfigure{\includegraphics[width=1\columnwidth]{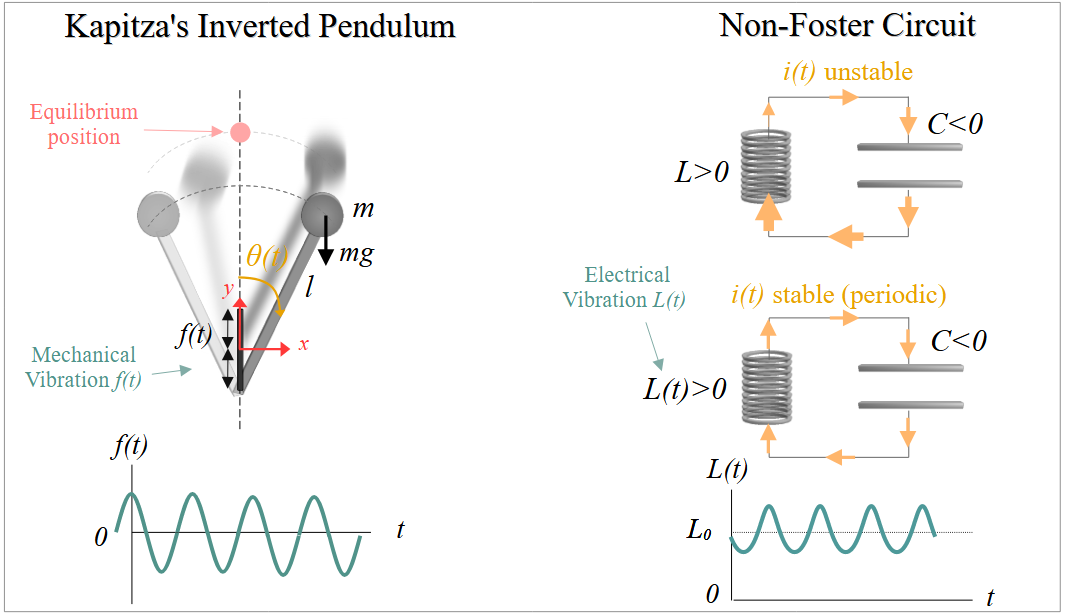}}
    \caption{Connection between Kapitza's inverted pendulum and time-modulated non-Foster circuits. Vibrational resonances appear in both systems, making them stable. } 
    \label{fig1}
\end{figure}

Let us consider the non-Foster $L(t)C$ resonator sketched in Figure~\ref{fig1}, formed by a time-invariant negative capacitor $C<0$ and a time-varying inductor $L(t)$. In the time-invariant capacitor, the relation between voltage $v_c(t)$ and current $i_c(t)$ is given by  $i_C(t) = C\, dv_C(t) / dt$. Similarly, the instantaneous relation between current $i_L(t)$ and voltage $v_L(t)$ in the time-varying inductor is given by $v_L(t) = d[L(t)\,i_L(t)] / dt$. Simple circuit analysis reveals that the differential equation that models the current in the non-Foster $L(t)C$ resonator takes the form
\begin{equation} \label{ode}
    i''(t) + p(t)\, i'(t) + q(t)\, i(t) = 0\, .
\end{equation}
 Prime ($'$) and double prime ($''$) notation indicates the first and second derivatives of the considered function. This equation is a second-order linear ordinary differential equation (ODE) with variable coefficients $p(t)$ and $q(t)$.  The variable coefficients are computed as
\begin{equation} \label{pq}
    p(t) \equiv  \frac{2 L'(t)}{L(t)}, \quad q(t) \equiv \frac{\frac{1}{C} + L''(t)}{L(t)}\, .
\end{equation}

The fact of having variable coefficients $p(t)$ and $q(t)$ makes it difficult to draw direct relationships and analogy between the $L(t)C$ circuit and Kapitza's inverted pendulum. Therefore, we now reduce the original ODE to a simpler equivalent one. Given  the change of variable
\begin{equation}
    i(t) = \phi(t)\, \mathrm{e}^{-\frac{1}{2} \int_0^t p(\tau) d\tau}\, ,
\end{equation}
a  second-order linear ODE such as eq.~\eqref{ode} can be transformed to the normal form \cite{Wynne1916}
\begin{equation} \label{Z}
    \phi''(t) + X(t)\, \phi(t) = 0\, ,
\end{equation}
where $X(t) \equiv q(t) -\frac{1}{2} p'(t) - \frac{1}{4} p^2(t)$.
For the variable coefficients $p(t)$ and $q(t)$ in eq.~\eqref{pq}, the term \linebreak $X(t) = 1 / [L(t)C]$ and the ODE for the new variable $\phi(t)$ takes the  form
\begin{equation} \label{z2}
    \phi''(t) + \frac{1}{L(t)C}\, \phi(t) = 0\, .
\end{equation}

In the case of the $L(t)C$ resonator, the variable coefficient $p(t)$ is written as $p(t) = 2 L'(t) / L(t)$, whose integral is $\int_0^t p(\tau) d\tau = 2 \ln [L(t)/L(0)]$. Thus, \linebreak  $\mathrm{e}^{-\frac{1}{2} \int_0^t p(\tau) d\tau} = L(0)/L(t)$ and
\begin{equation} \label{izL}
    i(t) = \phi(t) \frac{L(0)}{L(t)} \, ,
\end{equation}
\noindent for a generic modulation of the inductance $L(t)$ {and an initial state $L(0)$. It is worth noting that in the way eq.~\eqref{izL} is given, the term $\phi(t)$, which has the unit of current, essentially represents an equivalent magnetic flux (normalized by $L(0)$ associated to the inductor)}. 

To find the current $i(t)$ in eq.~\eqref{ode}, we can first solve for $\phi(t)$ in eq.~\eqref{z2} and then reapply the change of variable {$i(t) = \phi(t)L(0)/L(t)$}. This way of proceeding is more convenient, since it provides us with relevant information about the stability conditions of the system.  By inspecting eqs.~\eqref{z2} and~\eqref{izL}, it is clear that the current $i(t)$, so the non-Foster resonator, will be stable as long as $\phi(t)$ is stable and  {$L(0)/L(t)$} is bounded. Naturally, it is easier to set stability criteria for $\phi(t)$ in~\eqref{z2} than directly for eq.~\eqref{ode}. Moreover, the mathematical reduction of the original ODE now shows a direct physical connection between the non-Foster $L(t)C$ and the inverted pendulum. It can be shown (see the supplementary materials \cite{SuppMaterial}) that the ODE that describes the temporal evolution of the angle $\theta(t)$ in Kapitza's inverted pendulum is
\begin{equation} \label{theta_kapitza}
    \theta''(t) + \left[\alpha\Omega^2 \cos(\Omega t) - \omega_0^2 \right] \sin \left[\theta(t)\right] = 0\, ,
\end{equation}
In eq.~\eqref{theta_kapitza}, the parameter $\alpha \equiv \delta / l$ is a dimensionless constant that relates the amplitude of the added mechanical vibration $\delta$ and the fixed length of the pendulum $l$, $\Omega$ is the  angular frequency of the added mechanical vibration, and $\omega_0^2 \equiv g / l$ is the square of the natural oscillation frequency of the pendulum ($g\approx 9.81$ m/$s^2$ is the acceleration of gravity).  We note that the reduced ODE~\eqref{z2} in the $L(t)C$ resonator has the same exact form as the ODE~\eqref{theta_kapitza}  in Kapitza's inverted pendulum in the case of considering small angles $\theta(t)$. Note that in the small-angle approximation, $\sin [\theta(t)] \approx \theta(t)$. In fact,  it is possible to mimic the behavior of the inverted pendulum by selecting a proper modulation of the inductance $L(t)$. To do so, we have to enforce $\alpha\Omega^2 \cos(\Omega t) - \omega_0^2  = 1 /[L(t)C]$. Thus, the time modulation $L(t)$ that mimics the response of Kapitza's inverted pendulum is
\begin{equation} \label{L_kapitza}
    L(t) = \frac{1}{C\left[\alpha\Omega^2 \cos(\Omega t) - \omega_0^2 \right]}\, ,
\end{equation}
The associated current $i(t)$ is computed via eq.~\eqref{izL}:
{\begin{equation} \label{i_kapitza}
    i(t) = \phi(t)\, C \left[\alpha\Omega^2 \cos(\Omega t) - \omega_0^2 \right] L(0) \,. 
\end{equation}}
Inspection of eq.~\eqref{i_kapitza} reveals that the current $i(t)$ essentially represents an amplitude-modulated version of the temporal evolution of the angle in the inverted pendulum, as $\phi(t)$ effectively plays the role of $\theta(t)$.

Kapitza showed that the inverted pendulum can be stable in its upper vertical position as long as the frequency of the mechanical vibration $\Omega$ exceeds the limit value $\Omega_\mathrm{lim} = \sqrt{2} \omega_0 / \alpha$, which dictates that, no matter how small the relative amplitude of the applied vibration $\alpha$ is, we can always use a frequency $\Omega>\Omega_\mathrm{lim}$ high enough to make the inverted pendulum stable. Same rationale applies for the non-Foster $L(t)C$ resonator. When the applied time modulation of $L(t)$ is that of eq.~\eqref{L_kapitza}, the non-Foster resonator can become stable if the frequency of the \emph{electrical vibration} $\Omega$ exceeds the limit frequency $\Omega_\mathrm{lim}$. 

\begin{figure}[!t]
    \centering
\subfigure[]{\includegraphics[width=0.47\columnwidth]{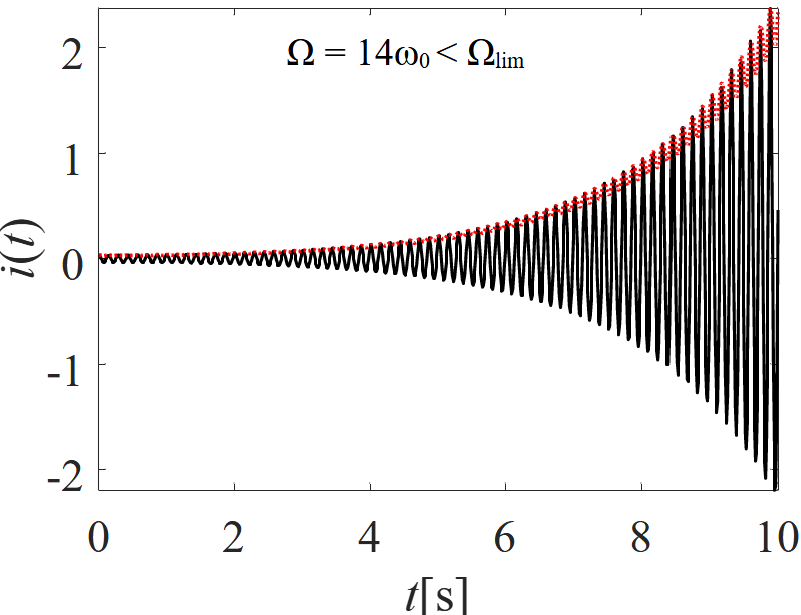}}
\hspace*{0.2cm}
\subfigure[]{\includegraphics[width=0.47\columnwidth]{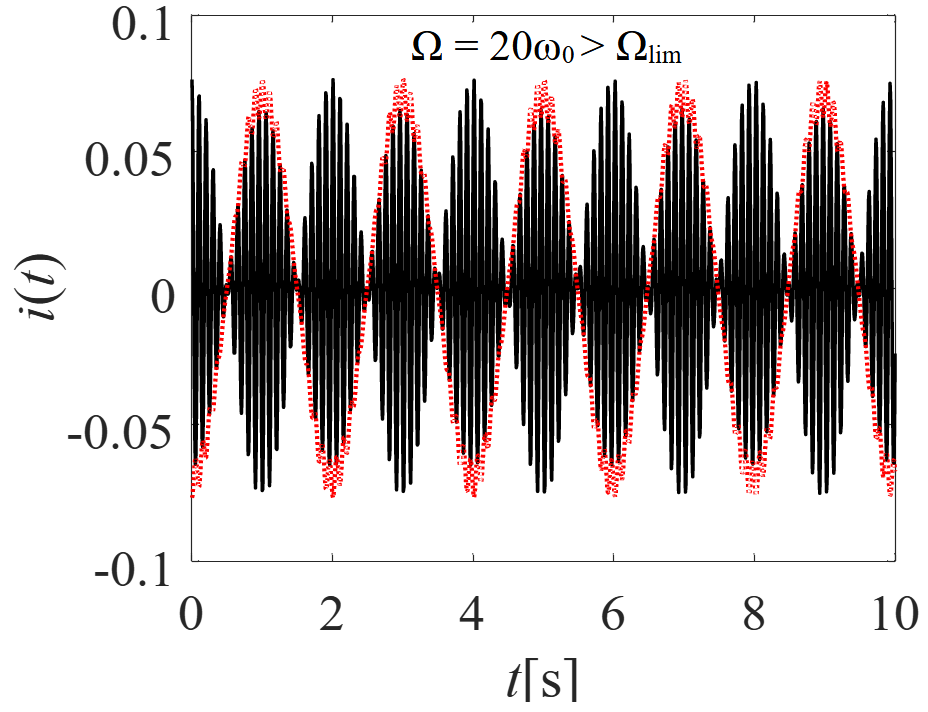}}
    \caption{Current $i(t)$ in the non-Foster $L(t)C$ resonator. The modulated $L(t)$ is the one that mimics the behavior of Kapitza's inverted pendulum [eq.~\eqref{L_kapitza}]. (a) Unstable case. (b) Stable case. The red dotted line represents a normalized version of $\phi(t)$. Parameters: $\alpha=0.1$, $C=-10^{-3}$ F.} 
    \label{fig2}
\end{figure}

Figure~\ref{fig2} illustrates the stabilization of the non-Foster circuit as a result of the applied modulation frequency $\Omega$. In order to compute the current in the circuit, we extract $\phi(t)$ by numerically solving the ODE~\eqref{z2} and then use eq.~\eqref{i_kapitza} to recover $i(t)$. The relative amplitude of the modulation is chosen to be $\alpha = 0.1$, a relatively small value. For this amplitude, the limit frequency that marks stability is $\Omega_\mathrm{lim} = \sqrt{2} \omega_0 / 0.1 \approx  14.14 \omega_0  $. Above  $\Omega_\mathrm{lim}$, the $L(t)C$ circuit will be stable. The considered initial conditions are $\phi(0) = \phi_0$ and $\phi'(0) = 0$, with $\phi_0 = -0.2$. These values translate into the following initial conditions for $i(t)$: $i(0) = \phi_0 C (\alpha \Omega^2 - \omega_0^2)$ and $i'(0) = 0$.

Figure~\ref{fig2}(a) shows an unstable case, with $\Omega$ less than the limit frequency. {In this scenario, the current grows exponentially over time. This is essentially due to the uncontrolled accumulation of external biasing energy needed to create the negative capacitance}. Figure~\ref{fig2}(b) shows a stable case, with $\Omega$ greater than the limit frequency.  The term $\phi(t)$ is essentially a scaled version of the envelope of the current $i(t)$ flowing through the non-Foster circuit. Naturally, this can also be inferred after inspecting eq.~\eqref{i_kapitza}. Red dotted line in Figures~\ref{fig2}(a) and (b) represent $\phi(t)$ after normalization. In addition, please note that, when the applied modulation of $L(t)$ is that of eq.~\eqref{L_kapitza}, the term $\phi(t)$ would also represent the angle $\theta(t)$ in a Kapitza's inverted pendulum. Thus, if $\phi(t)$ is stable, then the current $i(t)$ is stable too.

\section{\label{sec:Conc} Stabilization via positive  $L(t)$}

In the previous section, we have shown the physical correspondence between Kapitza's inverted pendulum and the non-Foster circuit. Moreover, we have illustrated that the non-Foster $L(t)C$  circuit can be stable if $L(t)$ is given by eq.~\eqref{L_kapitza} and the modulation frequency $\Omega$ exceeds Kapitza's limit value $\Omega_\mathrm{lim} = \sqrt{2}\omega_0/\alpha$. Unfortunately, the use of~\eqref{L_kapitza} presents two main impracticalities. The first one is that $|L(t)|\rightarrow \infty$ at some temporal instants $t$. Nonetheless, this impracticality is probably the least important, since it is possible to take sufficiently large, but finite, values of the inductance to approximate the asymptotic behavior of $L(t)$. The second and most important impracticality of using eq.~\eqref{L_kapitza} is that $L(t)$ becomes negative. Actually, $L(t)$ is negative approximately half of the modulation period and positive the other half. Not only do we need a time-varying inductor to stabilize the circuit, which is not easy to put into practice, but we also need it to alternate between positive and negative inductance values, which makes it effectively impossible. 

Now, the question that remains is: can we find a modulating $L(t)>0$ that avoids the use of negative and zero values? Eq.~\eqref{z2} leaves this possibility open. If it is possible to find a stable $\phi(t)$ for a positive-valued time-modulated inductor $L(t)$, then the current $i(t)$ (and thus the voltage) should be stable too. Finding via trial and error a non-negative expression for $L(t)$  that makes the non-Foster system stable  is really inefficient.  In this regard, the technique developed in \cite{Hrabar2020, Grigorii2023} can be helpful. The underlying idea there is that temporal modulations of a circuit element with positive values (e.g., a capacitor) allow to mimic the response of positive- or negative-valued time-invariant circuit elements.  A double-negative ($L_\mathrm{eq}<0$ and $C<0$) non-Foster resonator is naturally stable. Thus, if a positive time-varying inductor $L(t)>0$ behaves, equivalently, as a negative-valued time-invariant inductor $L_\mathrm{eq}<0$, then the time-varying non-Foster resonator will be stable.  

In order to ensure that the time-varying inductor $L(t)$ behaves exactly the same as an equivalent (negative-valued) time-invariant inductor $L_\mathrm{eq}$, voltages and current flowing through both terminals have to be identical ($d[L(t) i_L(t)]/dt = L_\mathrm{eq}\, d[ i_{L\mathrm{eq}}(t)]/dt$ and $i_L(t) = i_{L\mathrm{eq}}(t)$). Voltage and current equalities lead to the expression for $L(t)$: 
\begin{equation} \label{L_time}
    L(t) = L_\mathrm{eq} + \frac{c_1}{i_{L\mathrm{eq}}(t)}\, ,
\end{equation}
where $c_1$ is a generic constant of integration. The presence of the constant $c_1$ highlights the fact that there are infinite solutions for $L(t)$   that replicate the response of the time-invariant inductor $L_\mathrm{eq}$. 

In a conventional $L_\mathrm{eq}C$ resonator formed by a negative inductor $L_\mathrm{eq}$ and a negative capacitor $C$, the current $i_{L\mathrm{eq}}$ flowing through the inductor can be computed as $i_{L\mathrm{eq}} = -i_C = -C\, dv/dt$. In this scenario, the temporal evolution of the voltage 
across a capacitor $C$ initially charged with the conditions $v(0) = v_0$ and $v'(0) = 0$ will be of the form $v(t) = v_0 \cos(\omega_\mathrm{eq} t)$. The term  $\omega_\mathrm{eq} = 1 / \sqrt{L_\mathrm{eq}C}$ represents the natural oscillation frequency of the equivalent $L_\mathrm{eq}C$ resonator. Taking the derivative of $v(t)$, the current $i_{L\mathrm{eq}}(t)$ can be expressed as $i_{L\mathrm{eq}}(t) = +v_0 C\omega_\mathrm{eq} \sin(\omega_\mathrm{eq} t)$.

The former expression for the current $i_{L\mathrm{eq}}$ should be inserted into eq.~\eqref{L_time} to compute the expression of the time-varying inductor $L(t)$ that mimics the behavior of the negative $L_{\mathrm{eq}}$. However, the expression for $L(t)$ could still be giving negative values. Moreover, the current $i_{L\mathrm{eq}}(t)$ nulls at the instants $\omega_\mathrm{eq}t=n\pi$, with $n=0, 1, 2...$, leading to singularities in $L(t)$. Both inconveniences can be solved by connecting a DC current source $I$ in parallel with the $L(t)C$ and $L_\mathrm{eq}C$ resonators.  After connecting the DC current source, the current $i_{L\mathrm{eq}}$ is simply computed as $i_{L\mathrm{eq}}(t) = I-i_C(t) = I + v_0 C\omega_\mathrm{eq} \sin(\omega_\mathrm{eq} t)$. This leads to a time modulation of $L(t)$ of the form
\begin{equation} \label{L_time2}
    L(t) = L_\mathrm{eq} + \frac{c_1}{I + v_0 C\omega_\mathrm{eq} \sin(\omega_\mathrm{eq} t)}\, .
\end{equation}

By inspecting eq.~\eqref{L_time2}, it can be readily inferred that it is possible to select $c_1$ and a DC source $I$ that generate positive  values of $L(t)$ for all time instants, i.e., $L(t)>0$. This modulation is indeed more convenient from a practical perspective that the one mimicking the response of Kapitza's inverted pendulum [eq.~\eqref{L_kapitza}], which also gave stability but at the cost of requiring negative time-modulated inductance values. Moreover, the created stability in the non-Foster circuit via modulations of the form~\eqref{L_time2} can be explained based on the stability of another known system \cite{DAngelo1970}: the time-invariant $L_\mathrm{eq}C$ resonator with $L_\mathrm{eq}<0$ and $C<0$. The reader is referred to the Supplementary Material for further information \cite{SuppMaterial}.

The proposed positive time modulation of $L(t)$ in eq.~\eqref{L_time2} mimics the response of an equivalent negative $L_\mathrm{eq}$. Therefore, the global response of the non-Foster $L(t)C$ resonator connected to a DC current source is expected to be stable. In fact, eq.~\eqref{L_time2} fixes \emph{a priori} the conditions of stability for the non-Foster circuit. Provided that $L_\mathrm{eq}$ and $C$ are negative, voltage $v(t)$ and current $i(t)$ flowing through the time-varying inductor are periodic, of frequency $\omega_\mathrm{eq} = 1 / \sqrt{L_\mathrm{eq}C}$ and peak amplitudes $|v_0|$ and $|I - v_0C\omega_\mathrm{eq}|$, respectively. 

\begin{figure}[!t]
    \centering
\subfigure[]{\includegraphics[width=0.48\columnwidth]{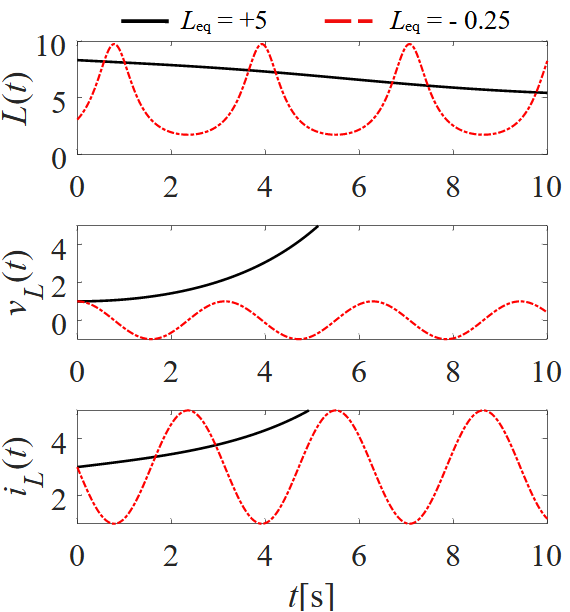}}
\hspace*{0.05cm}
\subfigure[]{\includegraphics[width=0.48\columnwidth]{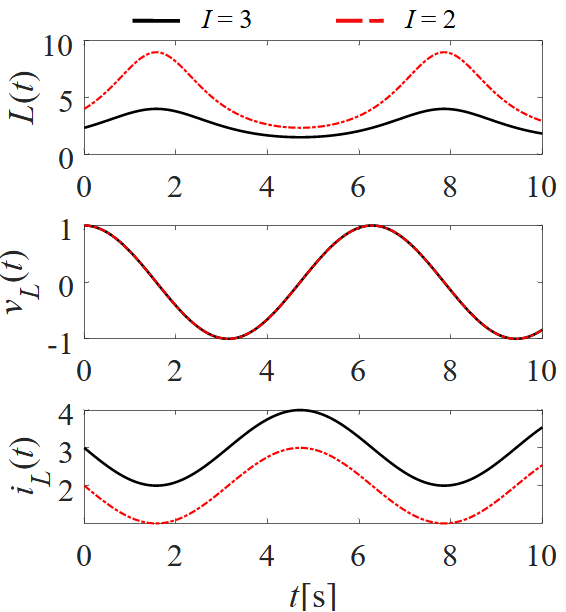}}
\caption{Stabilization of the non-Foster $L(t)C$ resonator with a positive-valued time modulated $L(t)>0$. (a) Effect of varying the equivalent inductance $L_\mathrm{eq}$ (for $c_1 = 10$, $I = 3$ A). (b) Effect of varying the DC current source (for $c_1 = 10$, $L_\mathrm{eq} = -1$ H).  In all cases, we have considered $C=-1$ F.  } 
    \label{fig3}
\end{figure}

Figure~\ref{fig3} illustrates a numerical example involving the use of the time modulation in eq.~\eqref{L_time2} and the stabilization of the non-Foster resonator. Top, middle and bottom panels represent the applied time modulated $L(t)$, the voltage in the time-varying inductor $v_L(t)$, and the current in the time-varying inductor $i_L(t)$, respectively. Figure~\ref{fig3}(a) shows two different scenarios for a $L(t)C$ resonator with $C=-1$ F. The selected time modulations, which follow the expression of eq.~\eqref{L_time2}, are positive in both cases, but lead to very different voltage and current curves. The black curve, where the time-varying modulation of $L(t)$ mimics the action of a positive-valued time-invariant equivalent inductance,  represents an unstable case. See how the voltage and current grow exponentially over time. On the other hand, the red curve represents a stable case, with $L(t)$ mimicking an equivalent negative-valued time-invariant $L_\mathrm{eq}$. In this case, the temporal variation is a periodic oscillation of frequency $\omega_\mathrm{eq}$. Figure~\ref{fig3}(a) clearly shows how the non-Foster $L(t)C$ resonator can be stabilized with a positive modulation $L(t)>0$. 

Figure~\ref{fig3}(b) illustrates the effect of the DC current source $I$ for a fixed equivalent inductance $L_\mathrm{eq} = -1$~H. As seen, the DC current source has no effect on the voltage $v_L(t)$. This is because the capacitor blocks its DC component. Conversely, the effect of $I$ is visible in the current $i_L(t)$. The DC current source simply raises the current in the inductor by a value of $I$. Additionally, although not explicitly shown, the constant $c_1$ has no effect on either voltage or current. From a practical perspective, the constant $c_1$ should be considered as a free parameter that, together with the DC current, helps to create a positive modulation for all time instants. 

\begin{figure}[!t]
    \centering
\subfigure{\includegraphics[width=0.8\columnwidth]{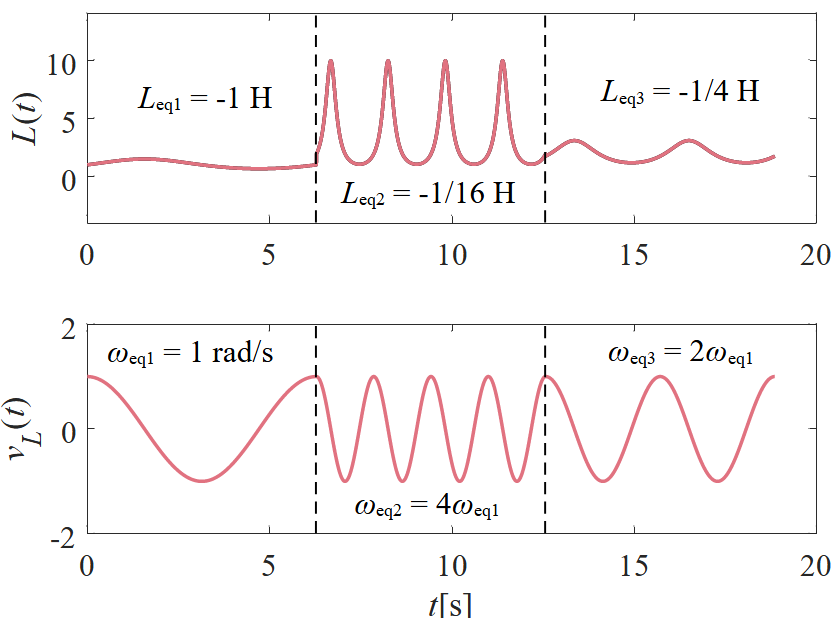}}
\caption{Dynamic reconfiguration of frequency in the non-Foster $L(t)C$ resonator. Parameters: $C=-1$ F, $I = 5$ A, $c_1 = 10$ Wb.  } 
    \label{fig4}
\end{figure}

Playing with the value of $L_{\rm eq}$ gives a route to dynamically tune the frequency $\omega_\mathrm{eq}$ of the already stabilized non-Foster circuit. This represents an interesting difference with respect to the conventional time-invariant $LC$ circuits, whose natural oscillation frequency is fixed  and cannot be changed for a given inductor and capacitor.  Thus, the $L(t)C$ circuit can be used as a frequency reconfiguration. Figure~\ref{fig4} illustrates this functionality. The value of $L_\mathrm{eq}$ is dynamically changed three times using three different temporal modulations of the positive-valued $L(t)$: from $L_\mathrm{eq1} = -1$ H  to $L_\mathrm{eq2} = -1/16$ H, and then to $L_\mathrm{eq3} = -1/4$ H. For a capacitor $C= -1$ F, the associated equivalent frequencies are $\omega_\mathrm{eq1} = 1$ rad/s, $\omega_\mathrm{eq2} = 4\omega_\mathrm{eq1}$, $\omega_\mathrm{eq3} = 2\omega_\mathrm{eq1}$, respectively. Figure~\ref{fig4}(a) shows the time-varying inductance $L(t)$ that creates the three equivalent inductances. Figure~\ref{fig4}(b) illustrates how the voltage in the inductor dynamically changes as a result of changing $L(t)$. Results show that frequency reconfiguration/tuning can be achieved with the non-Foster resonator. It should be noted that the stability of the results is also dependent on the discontinuities generated in the changes of $L(t)$ and the applied initial conditions. Moreover, it is expected that frequency tuning can also be achieved with the analogue Foster version of the resonator, with positive $C$ and a time modulation of $L(t)$ that mimics positive equivalent inductances.

\section{\label{sec:Conc} Conclusion}

In this paper, inspired by Kapitza's ideas, we have shown that non-Foster $L(t)C$ resonators with time-varying inductors and negative capacitors can be stabilized with the inclusion of temporal modulations.  We have found a connection between Kapitza's work on the inverted pendulum with a mechanically-vibrating base and time-modulated electric circuits. Eq.~\eqref{L_kapitza} provides a time modulation of $L(t)$ to create a one-to-one correspondence between the angle $\theta(t)$ in Kapitza's pendulum and the current $i(t)$ in the non-Foster $L(t)C$ circuit. We have described how $i(t)$ essentially represents an amplitude-modulated version of  $\theta(t)$. Unfortunately, the expression of $L(t)$ in eq.~\eqref{L_kapitza} involves negative values and is not convenient from a practical perspective. In order to find a  $L(t)>0$ that stabilizes the non-Foster circuit with $C<0$, we apply the technique introduced in \cite{Grigorii2023}. With the positive $L(t)>0$ given by eq.~\eqref{L_time2} (and the addition of a DC current source $I$), we mimic the behavior of an equivalent negative inductor $L_\mathrm{eq}<0$, thus stabilizing the non-Foster circuit. Moreover, how the oscillation frequency $\omega_\mathrm{eq}$ of the already stabilized circuit can be reconfigured in real time, giving a route for frequency tuning.  These results provide a new approach for stabilizing electric circuits through time modulations and suggest \textit{Vibrational Electromagnetics} as a foundation for applications in  {more complex circuits, transmission lines (concatenation of $LC$ stages), and generic space-time media.}

\begin{acknowledgments}
This work was supported in part by the Simons Collaboration on Extreme Wave Phenomena (grant SFI-MPS-EWP-00008530-04) and by the Ulla Tuominen Foundation to G.P.
\end{acknowledgments}

\section*{Data Availability}
The data that support the findings of this paper are
openly available \cite{Data}.

\bibliography{./ref}

\providecommand{\noopsort}[1]{}\providecommand{\singleletter}[1]{#1}%
\begin{thebibliography}{51}%
\makeatletter
\providecommand \@ifxundefined [1]{%
 \@ifx{#1\undefined}
}%
\providecommand \@ifnum [1]{%
 \ifnum #1\expandafter \@firstoftwo
 \else \expandafter \@secondoftwo
 \fi
}%
\providecommand \@ifx [1]{%
 \ifx #1\expandafter \@firstoftwo
 \else \expandafter \@secondoftwo
 \fi
}%
\providecommand \natexlab [1]{#1}%
\providecommand \enquote  [1]{``#1''}%
\providecommand \bibnamefont  [1]{#1}%
\providecommand \bibfnamefont [1]{#1}%
\providecommand \citenamefont [1]{#1}%
\providecommand \href@noop [0]{\@secondoftwo}%
\providecommand \href [0]{\begingroup \@sanitize@url \@href}%
\providecommand \@href[1]{\@@startlink{#1}\@@href}%
\providecommand \@@href[1]{\endgroup#1\@@endlink}%
\providecommand \@sanitize@url [0]{\catcode `\\12\catcode `\$12\catcode `\&12\catcode `\#12\catcode `\^12\catcode `\_12\catcode `\%12\relax}%
\providecommand \@@startlink[1]{}%
\providecommand \@@endlink[0]{}%
\providecommand \url  [0]{\begingroup\@sanitize@url \@url }%
\providecommand \@url [1]{\endgroup\@href {#1}{\urlprefix }}%
\providecommand \urlprefix  [0]{URL }%
\providecommand \Eprint [0]{\href }%
\providecommand \doibase [0]{https://doi.org/}%
\providecommand \selectlanguage [0]{\@gobble}%
\providecommand \bibinfo  [0]{\@secondoftwo}%
\providecommand \bibfield  [0]{\@secondoftwo}%
\providecommand \translation [1]{[#1]}%
\providecommand \BibitemOpen [0]{}%
\providecommand \bibitemStop [0]{}%
\providecommand \bibitemNoStop [0]{.\EOS\space}%
\providecommand \EOS [0]{\spacefactor3000\relax}%
\providecommand \BibitemShut  [1]{\csname bibitem#1\endcsname}%
\let\auto@bib@innerbib\@empty
\bibitem [{\citenamefont {Kapitza}(1951)}]{Kapitza1}%
  \BibitemOpen
  \bibfield  {author} {\bibinfo {author} {\bibfnamefont {P.~L.}\ \bibnamefont {Kapitza}},\ }\bibfield  {title} {\bibinfo {title} {Dynamic stability of the pendulum with vibrating suspension point (in russian)},\ }\href@noop {} {\bibfield  {journal} {\bibinfo  {journal} {Sov. Phys JETP}\ }\textbf {\bibinfo {volume} {21}},\ \bibinfo {pages} {588} (\bibinfo {year} {1951})}\BibitemShut {NoStop}%
\bibitem [{\citenamefont {Kapitza}(1965)}]{Kapitza2}%
  \BibitemOpen
  \bibfield  {author} {\bibinfo {author} {\bibfnamefont {P.~L.}\ \bibnamefont {Kapitza}},\ }\bibfield  {title} {\bibinfo {title} {Collected papers of {P. L. Kapitza}, {Volume} 2}\ }(\bibinfo  {publisher} {Pergamon, Oxford},\ \bibinfo {year} {1965})\BibitemShut {NoStop}%
\bibitem [{\citenamefont {Stephenson}(1908)}]{Stephenson1908}%
  \BibitemOpen
  \bibfield  {author} {\bibinfo {author} {\bibfnamefont {A.}~\bibnamefont {Stephenson}},\ }\bibfield  {title} {\bibinfo {title} {On induced stability},\ }\href@noop {} {\bibfield  {journal} {\bibinfo  {journal} {The London, Edinburgh, and Dublin Philosophical Magazine and Journal of Science}\ }\textbf {\bibinfo {volume} {15}},\ \bibinfo {pages} {233} (\bibinfo {year} {1908})}\BibitemShut {NoStop}%
\bibitem [{\citenamefont {Butikov}(2017)}]{Butikov2017KapitzaS}%
  \BibitemOpen
  \bibfield  {author} {\bibinfo {author} {\bibfnamefont {E.~I.}\ \bibnamefont {Butikov}},\ }\bibfield  {title} {\bibinfo {title} {Kapitza ’ s pendulum : A physically transparent simple treatment}\ }(\bibinfo {year} {2017})\BibitemShut {NoStop}%
\bibitem [{Sup()}]{SuppMaterial}%
  \BibitemOpen
  \href@noop {} {}\bibinfo {note} {See Supplemental Material at LINK TO BE INSERTED BY THE PUBLISHER}\BibitemShut {NoStop}%
\bibitem [{\citenamefont {Yang}\ \emph {et~al.}(2024)\citenamefont {Yang}, \citenamefont {Rajasekar},\ and\ \citenamefont {Sanjuán}}]{Yang2024}%
  \BibitemOpen
  \bibfield  {author} {\bibinfo {author} {\bibfnamefont {J.}~\bibnamefont {Yang}}, \bibinfo {author} {\bibfnamefont {S.}~\bibnamefont {Rajasekar}},\ and\ \bibinfo {author} {\bibfnamefont {M.~A.}\ \bibnamefont {Sanjuán}},\ }\bibfield  {title} {\bibinfo {title} {Vibrational resonance: A review},\ }\href@noop {} {\bibfield  {journal} {\bibinfo  {journal} {Physics Reports}\ }\textbf {\bibinfo {volume} {1067}},\ \bibinfo {pages} {1} (\bibinfo {year} {2024})}\BibitemShut {NoStop}%
\bibitem [{\citenamefont {Blekhman}(2000)}]{Blekhman2000}%
  \BibitemOpen
  \bibfield  {author} {\bibinfo {author} {\bibfnamefont {I.}~\bibnamefont {Blekhman}},\ }\href@noop {} {\emph {\bibinfo {title} {Vibrational Mechanics: Nonlinear Dynamic Effects, General Approach, Applications}}}\ (\bibinfo  {publisher} {World Scientific, Singapore},\ \bibinfo {year} {2000})\BibitemShut {NoStop}%
\bibitem [{\citenamefont {Blekhman}(2004)}]{Blekhman2004}%
  \BibitemOpen
  \bibfield  {author} {\bibinfo {author} {\bibfnamefont {I.}~\bibnamefont {Blekhman}},\ }\href@noop {} {\emph {\bibinfo {title} {Selected Topics in Vibrational Mechanics}}}\ (\bibinfo  {publisher} {World Scientific, Singapore},\ \bibinfo {year} {2004})\BibitemShut {NoStop}%
\bibitem [{\citenamefont {Landa}\ and\ \citenamefont {McClintock}()}]{Landa2000}%
  \BibitemOpen
  \bibfield  {author} {\bibinfo {author} {\bibfnamefont {P.~S.}\ \bibnamefont {Landa}}\ and\ \bibinfo {author} {\bibfnamefont {P.~V.~E.}\ \bibnamefont {McClintock}},\ }\bibfield  {title} {\bibinfo {title} {Vibrational resonance},\ }\href@noop {} {\bibfield  {journal} {\bibinfo  {journal} {J. Phys. A: Math. Gen.}\ }\textbf {\bibinfo {volume} {33}},\ \bibinfo {pages} {433}}\BibitemShut {NoStop}%
\bibitem [{\citenamefont {Jothimurugan}\ \emph {et~al.}(2013)\citenamefont {Jothimurugan}, \citenamefont {Thamilmaran}, \citenamefont {Rajasekar},\ and\ \citenamefont {Sanjuan}}]{Chua2013}%
  \BibitemOpen
  \bibfield  {author} {\bibinfo {author} {\bibfnamefont {R.}~\bibnamefont {Jothimurugan}}, \bibinfo {author} {\bibfnamefont {K.}~\bibnamefont {Thamilmaran}}, \bibinfo {author} {\bibfnamefont {S.}~\bibnamefont {Rajasekar}},\ and\ \bibinfo {author} {\bibfnamefont {M.~A.~F.}\ \bibnamefont {Sanjuan}},\ }\bibfield  {title} {\bibinfo {title} {Experimental evidence for vibrational resonance and enhanced signal transmission in {Chua's} circuit},\ }\href@noop {} {\bibfield  {journal} {\bibinfo  {journal} {International Journal of Bifurcation and Chaos}\ }\textbf {\bibinfo {volume} {23}},\ \bibinfo {pages} {1350189} (\bibinfo {year} {2013})}\BibitemShut {NoStop}%
\bibitem [{\citenamefont {Chizhevsky}(2021)}]{Chizhevsky2021}%
  \BibitemOpen
  \bibfield  {author} {\bibinfo {author} {\bibfnamefont {V.~N.}\ \bibnamefont {Chizhevsky}},\ }\bibfield  {title} {\bibinfo {title} {Amplification of optical signals in a bistable vertical-cavity surface-emitting laser by vibrational resonance},\ }\href@noop {} {\bibfield  {journal} {\bibinfo  {journal} {Phil. Trans. R. Soc. A.}\ }\textbf {\bibinfo {volume} {379}},\ \bibinfo {pages} {20200241} (\bibinfo {year} {2021})}\BibitemShut {NoStop}%
\bibitem [{\citenamefont {Murali}\ \emph {et~al.}(2021)\citenamefont {Murali}, \citenamefont {Rajasekar}, \citenamefont {Aravind}, \citenamefont {Kohar}, \citenamefont {Ditto},\ and\ \citenamefont {Sinha}}]{Murali2021}%
  \BibitemOpen
  \bibfield  {author} {\bibinfo {author} {\bibfnamefont {K.}~\bibnamefont {Murali}}, \bibinfo {author} {\bibfnamefont {S.}~\bibnamefont {Rajasekar}}, \bibinfo {author} {\bibfnamefont {M.~V.}\ \bibnamefont {Aravind}}, \bibinfo {author} {\bibfnamefont {V.}~\bibnamefont {Kohar}}, \bibinfo {author} {\bibfnamefont {W.~L.}\ \bibnamefont {Ditto}},\ and\ \bibinfo {author} {\bibfnamefont {S.}~\bibnamefont {Sinha}},\ }\bibfield  {title} {\bibinfo {title} {Construction of logic gates exploiting resonance phenomena in nonlinear systems},\ }\href@noop {} {\bibfield  {journal} {\bibinfo  {journal} {Phil. Trans. R. Soc. A.}\ }\textbf {\bibinfo {volume} {379}},\ \bibinfo {pages} {20200238} (\bibinfo {year} {2021})}\BibitemShut {NoStop}%
\bibitem [{\citenamefont {Zhang}\ \emph {et~al.}(2022)\citenamefont {Zhang}, \citenamefont {Jin}, \citenamefont {Xu},\ and\ \citenamefont {Yue}}]{Zhang2022}%
  \BibitemOpen
  \bibfield  {author} {\bibinfo {author} {\bibfnamefont {T.}~\bibnamefont {Zhang}}, \bibinfo {author} {\bibfnamefont {Y.}~\bibnamefont {Jin}}, \bibinfo {author} {\bibfnamefont {Y.}~\bibnamefont {Xu}},\ and\ \bibinfo {author} {\bibfnamefont {X.}~\bibnamefont {Yue}},\ }\bibfield  {title} {\bibinfo {title} {Dynamical response and vibrational resonance of a tri-stable energy harvester interfaced with a standard rectifier circuit},\ }\href@noop {} {\bibfield  {journal} {\bibinfo  {journal} {Chaos}\ }\textbf {\bibinfo {volume} {32}},\ \bibinfo {pages} {093150} (\bibinfo {year} {2022})}\BibitemShut {NoStop}%
\bibitem [{\citenamefont {Olusola}\ \emph {et~al.}(2020)\citenamefont {Olusola}, \citenamefont {Shomotun}, \citenamefont {Vincent},\ and\ \citenamefont {McClintock}}]{Olusola2020}%
  \BibitemOpen
  \bibfield  {author} {\bibinfo {author} {\bibfnamefont {O.~I.}\ \bibnamefont {Olusola}}, \bibinfo {author} {\bibfnamefont {O.~P.}\ \bibnamefont {Shomotun}}, \bibinfo {author} {\bibfnamefont {U.~E.}\ \bibnamefont {Vincent}},\ and\ \bibinfo {author} {\bibfnamefont {P.~V.~E.}\ \bibnamefont {McClintock}},\ }\bibfield  {title} {\bibinfo {title} {Quantum vibrational resonance in a dual-frequency-driven {Tietz-Hua} quantum well},\ }\href@noop {} {\bibfield  {journal} {\bibinfo  {journal} {Phys. Rev. E}\ }\textbf {\bibinfo {volume} {101}},\ \bibinfo {pages} {052216} (\bibinfo {year} {2020})}\BibitemShut {NoStop}%
\bibitem [{\citenamefont {Kuzmanovski}\ \emph {et~al.}(2024)\citenamefont {Kuzmanovski}, \citenamefont {Schmidt}, \citenamefont {Spaldin}, \citenamefont {R\o{}nnow}, \citenamefont {Aeppli},\ and\ \citenamefont {Balatsky}}]{KapitzaQuantum2024}%
  \BibitemOpen
  \bibfield  {author} {\bibinfo {author} {\bibfnamefont {D.}~\bibnamefont {Kuzmanovski}}, \bibinfo {author} {\bibfnamefont {J.}~\bibnamefont {Schmidt}}, \bibinfo {author} {\bibfnamefont {N.~A.}\ \bibnamefont {Spaldin}}, \bibinfo {author} {\bibfnamefont {H.~M.}\ \bibnamefont {R\o{}nnow}}, \bibinfo {author} {\bibfnamefont {G.}~\bibnamefont {Aeppli}},\ and\ \bibinfo {author} {\bibfnamefont {A.~V.}\ \bibnamefont {Balatsky}},\ }\bibfield  {title} {\bibinfo {title} {Kapitza stabilization of quantum critical order},\ }\href@noop {} {\bibfield  {journal} {\bibinfo  {journal} {Phys. Rev. X}\ }\textbf {\bibinfo {volume} {14}},\ \bibinfo {pages} {021016} (\bibinfo {year} {2024})}\BibitemShut {NoStop}%
\bibitem [{\citenamefont {Wang}\ \emph {et~al.}(2021{\natexlab{a}})\citenamefont {Wang}, \citenamefont {Yu}, \citenamefont {Ding}, \citenamefont {Li},\ and\ \citenamefont {Jia}}]{Wang2021}%
  \BibitemOpen
  \bibfield  {author} {\bibinfo {author} {\bibfnamefont {G.}~\bibnamefont {Wang}}, \bibinfo {author} {\bibfnamefont {D.}~\bibnamefont {Yu}}, \bibinfo {author} {\bibfnamefont {Q.}~\bibnamefont {Ding}}, \bibinfo {author} {\bibfnamefont {T.}~\bibnamefont {Li}},\ and\ \bibinfo {author} {\bibfnamefont {Y.}~\bibnamefont {Jia}},\ }\bibfield  {title} {\bibinfo {title} {Effects of electric field on multiple vibrational resonances in {Hindmarsh-Rose} neuronal systems},\ }\href@noop {} {\bibfield  {journal} {\bibinfo  {journal} {Chaos, Solitons \& Fractals}\ }\textbf {\bibinfo {volume} {150}},\ \bibinfo {pages} {111210} (\bibinfo {year} {2021}{\natexlab{a}})}\BibitemShut {NoStop}%
\bibitem [{\citenamefont {Rizza}\ and\ \citenamefont {Ciattoni}(2013)}]{Rizza2013}%
  \BibitemOpen
  \bibfield  {author} {\bibinfo {author} {\bibfnamefont {C.}~\bibnamefont {Rizza}}\ and\ \bibinfo {author} {\bibfnamefont {A.}~\bibnamefont {Ciattoni}},\ }\bibfield  {title} {\bibinfo {title} {Effective medium theory for kapitza stratified media: Diffractionless propagation},\ }\href@noop {} {\bibfield  {journal} {\bibinfo  {journal} {Phys. Rev. Lett.}\ }\textbf {\bibinfo {volume} {110}},\ \bibinfo {pages} {143901} (\bibinfo {year} {2013})}\BibitemShut {NoStop}%
\bibitem [{\citenamefont {Muniz}\ \emph {et~al.}(2019)\citenamefont {Muniz}, \citenamefont {Alberucci}, \citenamefont {Jisha}, \citenamefont {Monika}, \citenamefont {Nolte}, \citenamefont {Morandotti},\ and\ \citenamefont {Peschel}}]{Muniz19}%
  \BibitemOpen
  \bibfield  {author} {\bibinfo {author} {\bibfnamefont {A.~L.~M.}\ \bibnamefont {Muniz}}, \bibinfo {author} {\bibfnamefont {A.}~\bibnamefont {Alberucci}}, \bibinfo {author} {\bibfnamefont {C.~P.}\ \bibnamefont {Jisha}}, \bibinfo {author} {\bibfnamefont {M.}~\bibnamefont {Monika}}, \bibinfo {author} {\bibfnamefont {S.}~\bibnamefont {Nolte}}, \bibinfo {author} {\bibfnamefont {R.}~\bibnamefont {Morandotti}},\ and\ \bibinfo {author} {\bibfnamefont {U.}~\bibnamefont {Peschel}},\ }\bibfield  {title} {\bibinfo {title} {Kapitza light guiding in photonic mesh lattice},\ }\href@noop {} {\bibfield  {journal} {\bibinfo  {journal} {Opt. Lett.}\ }\textbf {\bibinfo {volume} {44}},\ \bibinfo {pages} {6013} (\bibinfo {year} {2019})}\BibitemShut {NoStop}%
\bibitem [{\citenamefont {Alberucci}\ \emph {et~al.}(2022)\citenamefont {Alberucci}, \citenamefont {Jisha}, \citenamefont {Monika}, \citenamefont {Peschel},\ and\ \citenamefont {Nolte}}]{Alberucci2022}%
  \BibitemOpen
  \bibfield  {author} {\bibinfo {author} {\bibfnamefont {A.}~\bibnamefont {Alberucci}}, \bibinfo {author} {\bibfnamefont {C.~P.}\ \bibnamefont {Jisha}}, \bibinfo {author} {\bibfnamefont {M.}~\bibnamefont {Monika}}, \bibinfo {author} {\bibfnamefont {U.}~\bibnamefont {Peschel}},\ and\ \bibinfo {author} {\bibfnamefont {S.}~\bibnamefont {Nolte}},\ }\bibfield  {title} {\bibinfo {title} {Wave manipulation via delay-engineered periodic potentials},\ }\href@noop {} {\bibfield  {journal} {\bibinfo  {journal} {Phys. Rev. Res.}\ }\textbf {\bibinfo {volume} {4}},\ \bibinfo {pages} {043162} (\bibinfo {year} {2022})}\BibitemShut {NoStop}%
\bibitem [{\citenamefont {Caloz}\ and\ \citenamefont {Deck-L{\'e}ger}(2019)}]{caloz2019spacetime}%
  \BibitemOpen
  \bibfield  {author} {\bibinfo {author} {\bibfnamefont {C.}~\bibnamefont {Caloz}}\ and\ \bibinfo {author} {\bibfnamefont {Z.-L.}\ \bibnamefont {Deck-L{\'e}ger}},\ }\bibfield  {title} {\bibinfo {title} {Spacetime metamaterials—part {I}: general concepts},\ }\href@noop {} {\bibfield  {journal} {\bibinfo  {journal} {IEEE Transactions on Antennas and Propagation}\ }\textbf {\bibinfo {volume} {68}},\ \bibinfo {pages} {1569} (\bibinfo {year} {2019})}\BibitemShut {NoStop}%
\bibitem [{\citenamefont {Galiffi}\ \emph {et~al.}(2022)\citenamefont {Galiffi}, \citenamefont {Tirole}, \citenamefont {Yin}, \citenamefont {Li}, \citenamefont {Vezzoli}, \citenamefont {Huidobro}, \citenamefont {Silveirinha}, \citenamefont {Sapienza}, \citenamefont {Al{\`u}},\ and\ \citenamefont {Pendry}}]{galiffi2022photonics}%
  \BibitemOpen
  \bibfield  {author} {\bibinfo {author} {\bibfnamefont {E.}~\bibnamefont {Galiffi}}, \bibinfo {author} {\bibfnamefont {R.}~\bibnamefont {Tirole}}, \bibinfo {author} {\bibfnamefont {S.}~\bibnamefont {Yin}}, \bibinfo {author} {\bibfnamefont {H.}~\bibnamefont {Li}}, \bibinfo {author} {\bibfnamefont {S.}~\bibnamefont {Vezzoli}}, \bibinfo {author} {\bibfnamefont {P.~A.}\ \bibnamefont {Huidobro}}, \bibinfo {author} {\bibfnamefont {M.~G.}\ \bibnamefont {Silveirinha}}, \bibinfo {author} {\bibfnamefont {R.}~\bibnamefont {Sapienza}}, \bibinfo {author} {\bibfnamefont {A.}~\bibnamefont {Al{\`u}}},\ and\ \bibinfo {author} {\bibfnamefont {J.}~\bibnamefont {Pendry}},\ }\bibfield  {title} {\bibinfo {title} {Photonics of time-varying media},\ }\href@noop {} {\bibfield  {journal} {\bibinfo  {journal} {Advanced Photonics}\ }\textbf {\bibinfo {volume} {4}},\ \bibinfo {pages} {014002} (\bibinfo {year} {2022})}\BibitemShut {NoStop}%
\bibitem [{\citenamefont {Engheta}(2023)}]{engheta2023four}%
  \BibitemOpen
  \bibfield  {author} {\bibinfo {author} {\bibfnamefont {N.}~\bibnamefont {Engheta}},\ }\bibfield  {title} {\bibinfo {title} {Four-dimensional optics using time-varying metamaterials},\ }\href@noop {} {\bibfield  {journal} {\bibinfo  {journal} {Science}\ }\textbf {\bibinfo {volume} {379}},\ \bibinfo {pages} {1190} (\bibinfo {year} {2023})}\BibitemShut {NoStop}%
\bibitem [{\citenamefont {Estep}\ \emph {et~al.}(2014)\citenamefont {Estep}, \citenamefont {Sounas}, \citenamefont {Soric},\ and\ \citenamefont {Alu}}]{estep2014magnetic}%
  \BibitemOpen
  \bibfield  {author} {\bibinfo {author} {\bibfnamefont {N.~A.}\ \bibnamefont {Estep}}, \bibinfo {author} {\bibfnamefont {D.~L.}\ \bibnamefont {Sounas}}, \bibinfo {author} {\bibfnamefont {J.}~\bibnamefont {Soric}},\ and\ \bibinfo {author} {\bibfnamefont {A.}~\bibnamefont {Alu}},\ }\bibfield  {title} {\bibinfo {title} {Magnetic-free non-reciprocity and isolation based on parametrically modulated coupled-resonator loops},\ }\href@noop {} {\bibfield  {journal} {\bibinfo  {journal} {Nature Physics}\ }\textbf {\bibinfo {volume} {10}},\ \bibinfo {pages} {923} (\bibinfo {year} {2014})}\BibitemShut {NoStop}%
\bibitem [{\citenamefont {Wu}\ \emph {et~al.}(2019)\citenamefont {Wu}, \citenamefont {Liu}, \citenamefont {Hickle}, \citenamefont {Peroulis}, \citenamefont {G{\'o}mez-D{\'\i}az},\ and\ \citenamefont {Melc{\'o}n}}]{wu2019isolating}%
  \BibitemOpen
  \bibfield  {author} {\bibinfo {author} {\bibfnamefont {X.}~\bibnamefont {Wu}}, \bibinfo {author} {\bibfnamefont {X.}~\bibnamefont {Liu}}, \bibinfo {author} {\bibfnamefont {M.~D.}\ \bibnamefont {Hickle}}, \bibinfo {author} {\bibfnamefont {D.}~\bibnamefont {Peroulis}}, \bibinfo {author} {\bibfnamefont {J.~S.}\ \bibnamefont {G{\'o}mez-D{\'\i}az}},\ and\ \bibinfo {author} {\bibfnamefont {A.~{\'A}.}\ \bibnamefont {Melc{\'o}n}},\ }\bibfield  {title} {\bibinfo {title} {Isolating bandpass filters using time-modulated resonators},\ }\href@noop {} {\bibfield  {journal} {\bibinfo  {journal} {IEEE Transactions on Microwave Theory and Techniques}\ }\textbf {\bibinfo {volume} {67}},\ \bibinfo {pages} {2331} (\bibinfo {year} {2019})}\BibitemShut {NoStop}%
\bibitem [{\citenamefont {Taravati}\ and\ \citenamefont {Eleftheriades}(2022)}]{taravati2022microwave}%
  \BibitemOpen
  \bibfield  {author} {\bibinfo {author} {\bibfnamefont {S.}~\bibnamefont {Taravati}}\ and\ \bibinfo {author} {\bibfnamefont {G.~V.}\ \bibnamefont {Eleftheriades}},\ }\bibfield  {title} {\bibinfo {title} {Microwave space-time-modulated metasurfaces},\ }\href@noop {} {\bibfield  {journal} {\bibinfo  {journal} {ACS Photonics}\ }\textbf {\bibinfo {volume} {9}},\ \bibinfo {pages} {305} (\bibinfo {year} {2022})}\BibitemShut {NoStop}%
\bibitem [{\citenamefont {Moreno-Rodríguez}\ \emph {et~al.}(2024)\citenamefont {Moreno-Rodríguez}, \citenamefont {Alex-Amor}, \citenamefont {Padilla}, \citenamefont {Valenzuela-Valdés},\ and\ \citenamefont {Molero}}]{Moreno23spacetime}%
  \BibitemOpen
  \bibfield  {author} {\bibinfo {author} {\bibfnamefont {S.}~\bibnamefont {Moreno-Rodríguez}}, \bibinfo {author} {\bibfnamefont {A.}~\bibnamefont {Alex-Amor}}, \bibinfo {author} {\bibfnamefont {P.}~\bibnamefont {Padilla}}, \bibinfo {author} {\bibfnamefont {J.~F.}\ \bibnamefont {Valenzuela-Valdés}},\ and\ \bibinfo {author} {\bibfnamefont {C.}~\bibnamefont {Molero}},\ }\bibfield  {title} {\bibinfo {title} {Space-time metallic metasurfaces for frequency conversion and beamforming},\ }\href@noop {} {\bibfield  {journal} {\bibinfo  {journal} {Phys. Rev. Appl.}\ }\textbf {\bibinfo {volume} {21}},\ \bibinfo {pages} {064018} (\bibinfo {year} {2024})}\BibitemShut {NoStop}%
\bibitem [{\citenamefont {Taravati}(2018)}]{taravati2018aperiodic}%
  \BibitemOpen
  \bibfield  {author} {\bibinfo {author} {\bibfnamefont {S.}~\bibnamefont {Taravati}},\ }\bibfield  {title} {\bibinfo {title} {Aperiodic space-time modulation for pure frequency mixing},\ }\href@noop {} {\bibfield  {journal} {\bibinfo  {journal} {Physical Review B}\ }\textbf {\bibinfo {volume} {97}},\ \bibinfo {pages} {115131} (\bibinfo {year} {2018})}\BibitemShut {NoStop}%
\bibitem [{\citenamefont {Li}\ \emph {et~al.}(2023)\citenamefont {Li}, \citenamefont {Ma}, \citenamefont {Bahrami}, \citenamefont {Deck-L\'eger},\ and\ \citenamefont {Caloz}}]{Li2023prism}%
  \BibitemOpen
  \bibfield  {author} {\bibinfo {author} {\bibfnamefont {Z.}~\bibnamefont {Li}}, \bibinfo {author} {\bibfnamefont {X.}~\bibnamefont {Ma}}, \bibinfo {author} {\bibfnamefont {A.}~\bibnamefont {Bahrami}}, \bibinfo {author} {\bibfnamefont {Z.-L.}\ \bibnamefont {Deck-L\'eger}},\ and\ \bibinfo {author} {\bibfnamefont {C.}~\bibnamefont {Caloz}},\ }\bibfield  {title} {\bibinfo {title} {Space-time fresnel prism},\ }\href@noop {} {\bibfield  {journal} {\bibinfo  {journal} {Phys. Rev. Appl.}\ }\textbf {\bibinfo {volume} {20}},\ \bibinfo {pages} {054029} (\bibinfo {year} {2023})}\BibitemShut {NoStop}%
\bibitem [{\citenamefont {Wang}\ \emph {et~al.}(2021{\natexlab{b}})\citenamefont {Wang}, \citenamefont {Asadchy}, \citenamefont {Fan},\ and\ \citenamefont {Tretyakov}}]{Wang2021power}%
  \BibitemOpen
  \bibfield  {author} {\bibinfo {author} {\bibfnamefont {X.}~\bibnamefont {Wang}}, \bibinfo {author} {\bibfnamefont {V.~S.}\ \bibnamefont {Asadchy}}, \bibinfo {author} {\bibfnamefont {S.}~\bibnamefont {Fan}},\ and\ \bibinfo {author} {\bibfnamefont {S.~A.}\ \bibnamefont {Tretyakov}},\ }\bibfield  {title} {\bibinfo {title} {Space–time metasurfaces for power combining of waves},\ }\href@noop {} {\bibfield  {journal} {\bibinfo  {journal} {ACS Photonics}\ }\textbf {\bibinfo {volume} {8}},\ \bibinfo {pages} {3034} (\bibinfo {year} {2021}{\natexlab{b}})}\BibitemShut {NoStop}%
\bibitem [{\citenamefont {Pacheco-Pe{\~n}a}\ \emph {et~al.}(2023)\citenamefont {Pacheco-Pe{\~n}a}, \citenamefont {Fink},\ and\ \citenamefont {Engheta}}]{pacheco2023temporal}%
  \BibitemOpen
  \bibfield  {author} {\bibinfo {author} {\bibfnamefont {V.}~\bibnamefont {Pacheco-Pe{\~n}a}}, \bibinfo {author} {\bibfnamefont {M.}~\bibnamefont {Fink}},\ and\ \bibinfo {author} {\bibfnamefont {N.}~\bibnamefont {Engheta}},\ }\bibfield  {title} {\bibinfo {title} {Temporal chirp, temporal lensing and temporal routing via space-time interfaces},\ }\href@noop {} {\bibfield  {journal} {\bibinfo  {journal} {arXiv preprint arXiv:2311.10855}\ } (\bibinfo {year} {2023})}\BibitemShut {NoStop}%
\bibitem [{\citenamefont {Hrabar}(2020)}]{Hrabar2020}%
  \BibitemOpen
  \bibfield  {author} {\bibinfo {author} {\bibfnamefont {S.}~\bibnamefont {Hrabar}},\ }\bibfield  {title} {\bibinfo {title} {Time-varying route to non-{Foster} elements},\ }in\ \href@noop {} {\emph {\bibinfo {booktitle} {2020 Fourteenth International Congress on Artificial Materials for Novel Wave Phenomena (Metamaterials)}}}\ (\bibinfo {year} {2020})\ pp.\ \bibinfo {pages} {108--109}\BibitemShut {NoStop}%
\bibitem [{\citenamefont {Hrabar}()}]{Hrabar2022}%
  \BibitemOpen
  \bibfield  {author} {\bibinfo {author} {\bibfnamefont {S.}~\bibnamefont {Hrabar}},\ }\bibfield  {title} {\bibinfo {title} {Time-varying versus non-{Foster} elements - similarities and differences},\ }in\ \href@noop {} {\emph {\bibinfo {booktitle} {2022 Sixteenth International Congress on Artificial Materials for Novel Wave Phenomena (Metamaterials)}}},\ pp.\ \bibinfo {pages} {X--199--X--201}\BibitemShut {NoStop}%
\bibitem [{\citenamefont {Ptitcyn}\ \emph {et~al.}(2022)\citenamefont {Ptitcyn}, \citenamefont {Mirmoosa}, \citenamefont {Hrabar},\ and\ \citenamefont {Tretyakov}}]{Grigorii2022_conference}%
  \BibitemOpen
  \bibfield  {author} {\bibinfo {author} {\bibfnamefont {G.}~\bibnamefont {Ptitcyn}}, \bibinfo {author} {\bibfnamefont {M.}~\bibnamefont {Mirmoosa}}, \bibinfo {author} {\bibfnamefont {S.}~\bibnamefont {Hrabar}},\ and\ \bibinfo {author} {\bibfnamefont {S.}~\bibnamefont {Tretyakov}},\ }\bibfield  {title} {\bibinfo {title} {Time-varying elements for realization of stable non-{Foster} circuits and metasurfaces},\ }in\ \href@noop {} {\emph {\bibinfo {booktitle} {16th International Congress on Artificial Materials for Novel Wave Phenomena - Metamaterials}}}\ (\bibinfo {year} {2022})\ pp.\ \bibinfo {pages} {1--3}\BibitemShut {NoStop}%
\bibitem [{\citenamefont {Ptitcyn}\ \emph {et~al.}(2023)\citenamefont {Ptitcyn}, \citenamefont {Mirmoosa}, \citenamefont {Hrabar},\ and\ \citenamefont {Tretyakov}}]{Grigorii2023}%
  \BibitemOpen
  \bibfield  {author} {\bibinfo {author} {\bibfnamefont {G.}~\bibnamefont {Ptitcyn}}, \bibinfo {author} {\bibfnamefont {M.}~\bibnamefont {Mirmoosa}}, \bibinfo {author} {\bibfnamefont {S.}~\bibnamefont {Hrabar}},\ and\ \bibinfo {author} {\bibfnamefont {S.}~\bibnamefont {Tretyakov}},\ }\bibfield  {title} {\bibinfo {title} {Time-modulated circuits and metasurfaces for emulating arbitrary transfer functions},\ }\href@noop {} {\bibfield  {journal} {\bibinfo  {journal} {Phys. Rev. Appl.}\ }\textbf {\bibinfo {volume} {20}},\ \bibinfo {pages} {014041} (\bibinfo {year} {2023})}\BibitemShut {NoStop}%
\bibitem [{\citenamefont {Foster}(1924)}]{Foster1924}%
  \BibitemOpen
  \bibfield  {author} {\bibinfo {author} {\bibfnamefont {R.~M.}\ \bibnamefont {Foster}},\ }\bibfield  {title} {\bibinfo {title} {A reactance theorem},\ }\href@noop {} {\bibfield  {journal} {\bibinfo  {journal} {Bell System Technical Journal}\ }\textbf {\bibinfo {volume} {3}},\ \bibinfo {pages} {259} (\bibinfo {year} {1924})}\BibitemShut {NoStop}%
\bibitem [{\citenamefont {Bode}(1945)}]{Bode1945}%
  \BibitemOpen
  \bibfield  {author} {\bibinfo {author} {\bibfnamefont {H.~W.}\ \bibnamefont {Bode}},\ }\href@noop {} {\emph {\bibinfo {title} {Network Analysis and Feedback Amplifier Design}}}\ (\bibinfo  {publisher} {Van Nostrand, New York},\ \bibinfo {year} {1945})\BibitemShut {NoStop}%
\bibitem [{\citenamefont {Fano}(1948)}]{Fano1948}%
  \BibitemOpen
  \bibfield  {author} {\bibinfo {author} {\bibfnamefont {R.~M.}\ \bibnamefont {Fano}},\ }\href@noop {} {\emph {\bibinfo {title} {Theoretical limitations on the broadband matching of arbitrary impedances (Technical Report No. 41)}}}\ (\bibinfo  {publisher} {Research Laboratory of Electronics, MIT},\ \bibinfo {year} {1948})\BibitemShut {NoStop}%
\bibitem [{\citenamefont {Rosa}\ and\ \citenamefont {Carlin}(1953)}]{Carlin1953}%
  \BibitemOpen
  \bibfield  {author} {\bibinfo {author} {\bibfnamefont {R.~L.}\ \bibnamefont {Rosa}}\ and\ \bibinfo {author} {\bibfnamefont {H.~J.}\ \bibnamefont {Carlin}},\ }\href@noop {} {\emph {\bibinfo {title} {A General Theory of Wideband Matching with Dissipative 4- Poles (Technical report)}}}\ (\bibinfo  {publisher} {Polytechnic Inst. Brooklyn, NY: Defense Tech. Info. Center, AD0002980},\ \bibinfo {year} {1953})\BibitemShut {NoStop}%
\bibitem [{\citenamefont {Sussman-Fort}\ and\ \citenamefont {Rudish}(2009)}]{Sussman2009}%
  \BibitemOpen
  \bibfield  {author} {\bibinfo {author} {\bibfnamefont {S.~E.}\ \bibnamefont {Sussman-Fort}}\ and\ \bibinfo {author} {\bibfnamefont {R.~M.}\ \bibnamefont {Rudish}},\ }\bibfield  {title} {\bibinfo {title} {Non-{Foster} impedance matching of electrically-small antennas},\ }\href@noop {} {\bibfield  {journal} {\bibinfo  {journal} {IEEE Transactions on Antennas and Propagation}\ }\textbf {\bibinfo {volume} {57}},\ \bibinfo {pages} {2230} (\bibinfo {year} {2009})}\BibitemShut {NoStop}%
\bibitem [{\citenamefont {Latour}(1923)}]{Latour1923}%
  \BibitemOpen
  \bibfield  {author} {\bibinfo {author} {\bibfnamefont {M.}~\bibnamefont {Latour}},\ }\bibfield  {title} {\bibinfo {title} {Negative impedance device},\ }\href@noop {} {\bibfield  {journal} {\bibinfo  {journal} {US Patent US1687253A}\ } (\bibinfo {year} {1923})}\BibitemShut {NoStop}%
\bibitem [{\citenamefont {Dolmage}(1927)}]{Dolmage1927}%
  \BibitemOpen
  \bibfield  {author} {\bibinfo {author} {\bibfnamefont {M.~M.}\ \bibnamefont {Dolmage}},\ }\bibfield  {title} {\bibinfo {title} {Negative resistance},\ }\href@noop {} {\bibfield  {journal} {\bibinfo  {journal} {US Patent US1863566A}\ } (\bibinfo {year} {1927})}\BibitemShut {NoStop}%
\bibitem [{\citenamefont {Mathes}(1927)}]{Mathes1927}%
  \BibitemOpen
  \bibfield  {author} {\bibinfo {author} {\bibfnamefont {R.~C.}\ \bibnamefont {Mathes}},\ }\bibfield  {title} {\bibinfo {title} {Negative impedance circuits},\ }\href@noop {} {\bibfield  {journal} {\bibinfo  {journal} {US Patent US1779382A}\ } (\bibinfo {year} {1927})}\BibitemShut {NoStop}%
\bibitem [{\citenamefont {Linvill}(1953)}]{Linvill1953}%
  \BibitemOpen
  \bibfield  {author} {\bibinfo {author} {\bibfnamefont {J.}~\bibnamefont {Linvill}},\ }\bibfield  {title} {\bibinfo {title} {Transistor negative-impedance converters},\ }\href@noop {} {\bibfield  {journal} {\bibinfo  {journal} {Proceedings of the IRE}\ }\textbf {\bibinfo {volume} {41}},\ \bibinfo {pages} {725} (\bibinfo {year} {1953})}\BibitemShut {NoStop}%
\bibitem [{\citenamefont {Ghadiri}\ and\ \citenamefont {Moez}(2014)}]{Ghadiri2014}%
  \BibitemOpen
  \bibfield  {author} {\bibinfo {author} {\bibfnamefont {A.}~\bibnamefont {Ghadiri}}\ and\ \bibinfo {author} {\bibfnamefont {K.}~\bibnamefont {Moez}},\ }\bibfield  {title} {\bibinfo {title} {Wideband active inductor and negative capacitance for broadband rf and microwave applications},\ }\href@noop {} {\bibfield  {journal} {\bibinfo  {journal} {IEEE Transactions on Components, Packaging and Manufacturing Technology}\ }\textbf {\bibinfo {volume} {4}},\ \bibinfo {pages} {1808} (\bibinfo {year} {2014})}\BibitemShut {NoStop}%
\bibitem [{\citenamefont {Covington}\ \emph {et~al.}(2014)\citenamefont {Covington}, \citenamefont {Smith}, \citenamefont {Kshatri}, \citenamefont {Shehan}, \citenamefont {Weldon},\ and\ \citenamefont {Adams}}]{Covington2014}%
  \BibitemOpen
  \bibfield  {author} {\bibinfo {author} {\bibfnamefont {J.~M.~C.}\ \bibnamefont {Covington}}, \bibinfo {author} {\bibfnamefont {K.~L.}\ \bibnamefont {Smith}}, \bibinfo {author} {\bibfnamefont {V.~S.}\ \bibnamefont {Kshatri}}, \bibinfo {author} {\bibfnamefont {J.~W.}\ \bibnamefont {Shehan}}, \bibinfo {author} {\bibfnamefont {T.~P.}\ \bibnamefont {Weldon}},\ and\ \bibinfo {author} {\bibfnamefont {R.~S.}\ \bibnamefont {Adams}},\ }\bibfield  {title} {\bibinfo {title} {A cross-coupled cmos negative capacitor for wideband metamaterial applications},\ }in\ \href@noop {} {\emph {\bibinfo {booktitle} {IEEE SOUTHEASTCON 2014}}}\ (\bibinfo {year} {2014})\ pp.\ \bibinfo {pages} {1--5}\BibitemShut {NoStop}%
\bibitem [{\citenamefont {Brownlie}(1966)}]{Brownlie1966}%
  \BibitemOpen
  \bibfield  {author} {\bibinfo {author} {\bibfnamefont {J.}~\bibnamefont {Brownlie}},\ }\bibfield  {title} {\bibinfo {title} {On the stability properties of a negative impedance converter},\ }\href@noop {} {\bibfield  {journal} {\bibinfo  {journal} {IEEE Transactions on Circuit Theory}\ }\textbf {\bibinfo {volume} {13}},\ \bibinfo {pages} {98} (\bibinfo {year} {1966})}\BibitemShut {NoStop}%
\bibitem [{\citenamefont {Stearns}(2012)}]{Stearns2012}%
  \BibitemOpen
  \bibfield  {author} {\bibinfo {author} {\bibfnamefont {S.~D.}\ \bibnamefont {Stearns}},\ }\bibfield  {title} {\bibinfo {title} {Incorrect stability criteria for non-foster circuits},\ }in\ \href@noop {} {\emph {\bibinfo {booktitle} {Proceedings of the 2012 IEEE International Symposium on Antennas and Propagation}}}\ (\bibinfo {year} {2012})\BibitemShut {NoStop}%
\bibitem [{\citenamefont {Tofigh}\ and\ \citenamefont {Ziolkowski}(2020)}]{Tofigh2020}%
  \BibitemOpen
  \bibfield  {author} {\bibinfo {author} {\bibfnamefont {F.}~\bibnamefont {Tofigh}}\ and\ \bibinfo {author} {\bibfnamefont {R.~W.}\ \bibnamefont {Ziolkowski}},\ }\bibfield  {title} {\bibinfo {title} {A stable floating non-foster circuit},\ }in\ \href@noop {} {\emph {\bibinfo {booktitle} {2020 4th Australian Microwave Symposium (AMS)}}}\ (\bibinfo {year} {2020})\ pp.\ \bibinfo {pages} {1--2}\BibitemShut {NoStop}%
\bibitem [{\citenamefont {Wynne}\ and\ \citenamefont {Spraragen}(1916)}]{Wynne1916}%
  \BibitemOpen
  \bibfield  {author} {\bibinfo {author} {\bibfnamefont {W.~E.}\ \bibnamefont {Wynne}}\ and\ \bibinfo {author} {\bibfnamefont {W.}~\bibnamefont {Spraragen}},\ }\href@noop {} {\emph {\bibinfo {title} {Handbook of Engineering Mathematics}}}\ (\bibinfo  {publisher} {New York, D. Van Nostrand company},\ \bibinfo {year} {1916})\BibitemShut {NoStop}%
\bibitem [{\citenamefont {D'Angelo}(1970)}]{DAngelo1970}%
  \BibitemOpen
  \bibfield  {author} {\bibinfo {author} {\bibfnamefont {H.}~\bibnamefont {D'Angelo}},\ }\href@noop {} {\emph {\bibinfo {title} {Linear Time-Varying Systems: Analysis and Synthesis}}}\ (\bibinfo  {publisher} {Allyn and Bacon, Inc., 470 Atlantic Avenue, Boston, USA},\ \bibinfo {year} {1970})\BibitemShut {NoStop}%
\bibitem [{Dat()}]{Data}%
  \BibitemOpen
  \href@noop {} {}\bibinfo {note} {The DOI to access the data and codes \url{https://doi.org/10.5281/zenodo.15839313}}\BibitemShut {NoStop}%
\end{thebibliography}%



\setcounter{equation}{0}
\setcounter{figure}{0}
\setcounter{table}{0}
\setcounter{page}{1}
\makeatletter
\renewcommand{\theequation}{S\arabic{equation}}
\renewcommand{\thefigure}{S\arabic{figure}}

\pagebreak

\widetext
\begin{center}
\textrm{\huge Supplementary Material}
\end{center}

\vspace{0.2cm}

\begin{center}
\textrm{\Large Kapitza-Inspired Stabilization of Non-Foster Circuits via Time Modulations}
\end{center}

\begin{center}
\textrm{\large Antonio Alex-Amor, Grigorii Ptitcyn, Nader Engheta}
\end{center}

\begin{center}
\textit{\large University of Pennsylvania, Department of Electrical and Systems Engineering, \\ Philadelphia, Pennsylvania 19104, United States}
\end{center}


\section*{Kapitza's Inverted Pendulum}

\subsection*{Equations}

Let us consider the Kapitza's inverted pendulum shown in Figure 1 \cite{Kapitza1, Kapitza2}. The center pivot is driven to oscillate up and down, according to the function 
$f(t)$. The position of the point mass $m$ in the Cartesian coordinate system is described by the coordinates
\begin{equation}
    x(t) = l \sin(\theta)\, ,
\end{equation}

\begin{equation}
    y(t) = l \cos(\theta) + f(t)\, .
\end{equation}

When $f(t) = \delta \cos(\Omega t)$, the Lagrangian of this lossless system is
\begin{equation}
    L = \frac{1}{2}m \left[l^2 (\theta')^2 + 2l\delta \Omega \sin(\Omega t) \sin(\theta)\, \theta' + \delta^2 \Omega^2 \sin^2(\Omega t)\right] - mg \left[ l\cos(\theta) + \delta \cos(\Omega t)\right]\, ,
\end{equation}
where the first addend represents the kinetic energy, and the second addend represents the potential energy (negatively signed) of the inverted pendulum.  The superscript (') represents the first derivative with respect to time.  

The application of the Euler-Lagrange equations, $d(\partial L / \partial \theta')/dt - \partial L / \partial \theta = 0$, leads to a second-order nonlinear ordinary differential equation that describes the evolution of the angle $\theta$ as a function of time:
\begin{equation}
    \theta''(t) + \sin\left[\theta(t)\right]\left[\alpha\Omega^2 \cos(\Omega t) - \omega_0^2 \right] = 0\, ,
\end{equation}
with $\alpha \equiv \delta / l$  and $\omega_0^2 \equiv g / l$ being a dimensionless constant and the square of the natural oscillation frequency, respectively. Note that in the case that the pendulum is not driven ($\delta = 0$), the former ODE reduces to the one of the conventional simple pendulum.  

\subsection*{Slow and Fast Components}

The governing ODE for the pendulum has a difficult analytical treatment \cite{Kapitza1, Kapitza2}. Kapitza, based on the idea of having a high-frequency vibrating component of different characteristics from the natural oscillation frequency of the pendulum, decided to split the angle $\theta$ into two terms: the slow component $\theta_s(t)$ and the fast component $\theta_f(t)$. The slow component is of high amplitude and low frequency, while the fast component is of small amplitude and high frequency. Thus, the ODE can be rewritten as
\begin{equation} \label{ode_pendulum}
\theta_s''(t) +\theta_f''(t)  +  \sin\left(\theta_s(t) + \theta_f(t)\right)\left[\alpha\Omega^2 \cos(\Omega t) - \omega_0^2 \right] = 0\, .
\end{equation}
From the fast component's perspective, the slow component is practically constant over the fast period $T=2\pi / \Omega$. Additionally, $|\theta_f(t)|\ll |\theta_s(t)|$. Under this assumptions, eq. \eqref{ode_pendulum} reduces to the following equation for the ODE of the fast component:
\begin{equation} \label{pendulum_fast}
    \theta_f''(t) + \alpha \Omega^2 \cos(\Omega t) \sin(\theta_s) \approx 0\,
\end{equation}
whose solution will be of the form
\begin{equation}
    \theta_f(t) \approx \alpha \cos(\Omega t) \sin(\theta_s)\, .
\end{equation}
As this equation shows, the fast component is a sort of small-amplitude version of the fast component \linebreak ($\theta_f \propto \alpha \sin(\theta_s)$) that vibrates very fast at a frequency $\Omega$.

In order to extract the ODE for the slow component, eq. \eqref{ode_pendulum} can be time-averaged over the fast period $T = 2\pi/\Omega$ with the information about $\theta_f(t)$ that it has been already extracted. This leads to
\begin{equation} \label{ode_pendulum_slow}
    \theta_s''(t) + \sin\left(\theta_s(t)\right) \left[ \frac{\alpha^2 \Omega^2}{2} \cos(\theta_s(t)) -\omega_0^2\right] \approx 0\, .
\end{equation}
Eq. \eqref{ode_pendulum_slow} gives us information about the temporal evolution of the slow component $\theta_s$. With information about the slow and fast components, the whole solution, $\theta(t) = \theta_s(t) + \theta_f(t)$, can be reconstructed. Although eq. \eqref{ode_pendulum_slow} cannot be easily treated analytically, it can be further manipulated in order to work directly with potentials and thus derive information about the stable positions, $\theta = \{0,\pi\}$, in
the inverted pendulum. 

\section*{Stability Theorem for $\phi(t)$}

Theorem 8.5 in \cite{DAngelo1970} gives guidance on the stability of one system based on the known stability of another. The mentioned theorem states that the solutions of linear time-varying system of the form 
\begin{equation}
    \mathbf{\Phi}'(t) + \big[X_0 + \Delta \mathbf{X}(t) \big]\mathbf{\Phi}(t) = \mathbf{0}
\end{equation}
are stable provided that the solutions of
\begin{equation}
    \mathbf{\Phi}'(t) + X_0\mathbf{\Phi}(t) = \mathbf{0}
\end{equation}
are stable too and that $\int_0^\infty || \Delta \mathbf{X}(t)||\, dt<\infty $, where $X_0$ is a constant.  

The stability of eq. (5), 
$$
\phi''(t) + \frac{1}{L(t)C}\, \phi(t) = 0\, ,
$$
and, thus, of the whole non-Foster circuit, for a modulation of the form  $L(t) = L_\mathrm{eq} + c_1 / i(t)$ can be proven with the application of Theorem 8.5. This is because the modulation $L(t)$, which is formed by the addition of a constant $L_\mathrm{eq}$ and a time-varying term $c_1 / i(t)$, resembles the form $X_0 + \Delta \mathbf{X}(t)$.  

We know that the solutions of
\begin{equation} \label{z_static}
    \phi''(t) + \frac{1}{L_\mathrm{eq}C}\, \phi(t) = 0
\end{equation}
are stable if the constant term $1/(L_\mathrm{eq}C)$ is positive. For the case considered in the manuscript, $C<0$ and $L_\mathrm{eq}<0$, the term $1/(L_\mathrm{eq}C)>0$, thus the solutions $\phi(t)$ in eq. \eqref{z_static} are stable. Therefore, the application of Theorem 8.5 in \cite{DAngelo1970} states that the non-Foster circuit is stable, if the applied modulation is of the form $L(t) = L_\mathrm{eq} + c_1 / i(t)$, with $L_\mathrm{eq}<0$ and $C<0$. This holds as long as the time-varying term $c_1 / i(t)$ does not lead to a divergent improper integral ($\int_0^\infty || \Delta \mathbf{X}(t)||\, dt<\infty $). i.e., as long as $i(t)\neq 0\, \forall t$ and $i(t)$ is bounded and decaying over time. Currents $i(t)$ of the form $i(t) = [I + i_0\sin(\omega t)] \mathrm{e}^{-at} \neq 0$ fulfill this condition. This selection of $i(t)$ is essentially similar to the one chosen in the manuscript, since it is possible to select an attenuation factor $a$ small enough ($a\rightarrow 0^+$) so that the decay is not appreciable over time and the non-Foster system is  stable.






\end{document}